\documentstyle[12pt]{article}
\begin{document}
\pagestyle{plain}
\setcounter{page}{1}
\newcommand{\be}{\begin{equation}}
\newcommand{\ee}{\end{equation}}
\newcommand{\bea}{\begin{eqnarray}}
\newcommand{\eea}{\end{eqnarray}}
\renewcommand{\theequation}{\arabic{section}.\arabic{equation}}
\newcommand{\nono}{\nonumber}
\newcommand{\tU}{\tilde{U}}
\newcommand{\Tr}{{\rm Tr}}
\newcommand{\tr}{{\rm tr}}
\newcommand{\ket}[1]{|#1\rangle}
\newcommand{\sta}[1]{|#1)}

\setcounter{section}{0}
\setcounter{subsection}{0}
\def\thefootnote{\fnsymbol{footnote}}
\begin{titlepage}


\begin{flushright}
UT-Komaba/99-3 \\
January 1999 \\
{\tt hep-th/9902004 }
\end{flushright}
\vspace{1cm}
\begin{center}
\huge
World Volume Noncommutativity \\
versus \\
Target Space Noncommutativity
\end{center}
\vspace{1cm}
\normalsize
\begin{center}
{\sc Mitsuhiro Kato} 
\footnote{
e-mail address:\ \  {\tt kato@hep1.c.u-tokyo.ac.jp}}
{\sc and Tsunehide Kuroki}
\footnote{
e-mail address:\ \  {\tt kuroki@hep1.c.u-tokyo.ac.jp}}

\vspace{0.3cm}
{\it Institute of Physics, University of Tokyo\\
Komaba, Meguro-ku, 153 Tokyo}
\end{center}
\vspace{7mm}
\begin{center}
{\large Abstract}
\end{center}
\noindent
It is known that the noncommutativity of D-brane coordinate is responsible for
describing the higher-dimensional D-branes in terms of more fundamental ones
such as D-particles or D-instantons, while considering a noncommutative torus 
as a target space is conjectured to be equivalent to introducing the background
antisymmetric tensor field in matrix models.
In the present paper we clarify the dual nature of both descriptions. Namely
the noncommutativity of conjugate momenta of the D-brane coordinates realizes
the target space structure, whereas noncommutativity of the coordinates
themselves realizes world volume structure. We explicitly construct a boundary
state for the Dirichlet boundary condition where the string boundary is adhered
to the D-brane on the noncommutative torus. There are non-trivial relations
between the parameters appeared in the algebra of the coordinates and that of
the momenta.\vspace{7mm}
\end{titlepage}


\newpage
\section{Introduction}
Recent developments in nonperturbative string theory have revealed 
various kinds of noncommutative nature. One of such noncommutativity 
is that the space-time coordinates of $N$ D-branes are promoted to 
non-commuting $N\times N$ matrices \cite{boundstate}.  In particular, 
this description of coordinates of D-particles exhibits the noncommutative 
nature at the sub-string scale \cite{shortdist} where conventional 
descriptions of space-time by the differential geometry cease to make sense. 
These facts suggest that noncommutative geometry (NCG) may play an important 
role as a mathematical tool in the nonperturbative formulation of string 
theory. The central idea of NCG is, using the equivalence of a manifold 
and the $C^*$-algebra of functions defined on it, to reformulate the 
geometry of manifold in terms of the commutative $C^*$-algebra and then 
to generalize the results to the case of a noncommutative algebra. 
Thus in the spirit of NCG, the world volumes of D-branes should realize NCG
because the coordinate functions on the world volumes of D-branes are 
non-commuting matrices. In fact, the size $N$ of the non-commuting matrices 
originates not from the target space structure, but from the number of the 
world volumes of D-branes. Henceforth we will refer to this noncommutative 
nature associated to the world volumes of D-branes as 
{\it world volume noncommutativity} which is characterized  
by the noncommutativity of the matrices of the embedding coordinates 
of D-branes.
Especially for the classical solutions describing D-brane with dimensions
higher than the original one, say D-instanton, we need to set the following
non-trivial commutation relations for the D-instanton coordinates
\be
[q^i, q^j]=-if^{ij},
\label{wvnc}
\ee
whereas the commuting case corresponds merely to multiple D-instantons
configuration.

Another kind of noncommutative nature in string theory has emerged in the 
context of the toroidal compactification of Matrix theory 
\cite{BFSS}.  It is well known that Matrix theory compactified on a torus is 
obtained as supersymmetric Yang-Mills (SYM) theory on the dual torus
\cite{tetal}. However, if we consider the compactification on a torus 
$T^d$ for $d\geq 2$, we have an additional moduli arising from the 3-form 
tensor field in the low energy 11-dimensional supergravity. 
In fact, it was proposed that Matrix theory compactified on a torus 
with this constant background 3-form field is described by SYM theory 
on the dual noncommutative torus \cite{cds}. Moreover, it is shown that 
when compactified on $T^2$ with a constant Neveu-Schwarz 2-form field 
$B_{ij}$, D-brane world volume theory is also naturally described 
by gauge theory on the noncommutative torus and that $\theta$-parameter 
which characterizes the noncommutativity of the torus should be identified 
with $B_{ij}$ \cite{dh}. Evidences and generalizations of these proposals 
have been studied in many papers \cite{howu1}-\cite{kim}. 
In the following, we will call such noncommutative nature characterized 
by $\theta\sim B_{ij}$ {\it target space noncommutativity} 
because it originates from the noncommutativity of the target space torus 
on which the D-brane world volume is compactified. 
     
Since both noncommutative natures have been revealed as nonperturbative 
aspects of string theory, it should be important from the point of view of 
nonperturbative formulation of string theory to analyze them, in particular, 
to clarify their relationship. This is the problem we address in this paper.  
We will show that they are in some sense dual to each other and obey a certain 
simple relation by considering a boundary state which describes a configuration
of D-instantons on the (non)commutative two-torus.
We will only consider the bosonic string in order to make our argument
as simple as possible, but the generalization to the supersymmetric case
is straightforward.
In our analysis, one can study how moduli of a D-string made 
of D-instantons are encoded in the noncommutativities in the original 
D-instanton configuration. 
{}From the point of view of D-instantons as the fundamental degrees of freedom 
like \cite{IKKT,yoneya,nb,topmm}, it is important to interpret the various 
moduli of the higher branes made of D-instantons and symmetries between them 
as degrees of freedom present in the D-instanton configuration, partly done 
in \cite{ishi}. Thus our results will serve as a key ingredient to seek 
the constructive definition of string theory by regarding D-instantons 
as the constituent.      
 
The organization of this paper is as follows. In section 2, we present a 
prescription for the compactification on the noncommutative torus 
$T^2_{\theta}$. In section 3, we construct a boundary state corresponding
to D-instantons on $T^2_{\theta}$ and show that in this realization 
the ``momentum'' operators of the D-instantons have noncommutative nature 
characterized by $-\theta$. Thus we manifest that the world volume 
noncommutativity and the target space noncommutativity are dual to each other. 
In section 4, a configuration of D-instantons whose coordinates are given 
by non-commuting matrices is studied on $T^2$ with the constant $B_{ij}$ flux. 
Following the boundary state analysis done in \cite{ishi}, it is found 
that the world volume and the target space noncommutativity are related 
through a simple equation which is nothing but the consistency condition for 
a twisted bundle on $T^2_{\theta}$. Implications of this result are also 
given. Section 5 is devoted to the discussions.

\section{Compactification on Noncommutative Two-Torus}
\setcounter{equation}{0}

In this section, following the prescriptions given in 
\cite{BFSS, tetal, howu2}, we present a formulation of the compactification 
on a noncommutative two-torus. 

We begin by recalling the usual compactification on $T^2$: 
$$
X^i \in {\bf R}^2/2\pi \Lambda^2,~~~\Lambda^2\simeq {\bf Z}^2,~~~
p_i\in (\Lambda^2)^*
$$
\be
\rightarrow\exp ip_i(X^i+2\pi L^i)=\exp ip_iX^i~~~\mbox{for}
{}~{}^{\forall}L^i\in \Lambda^2. 
\label{single}
\ee

Generalizing this, we define the compactification on the noncommutative 
two-torus $T^2_{\theta}$ as follows. First we introduce the projective 
representation of ${\bf Z}^2$ which is labeled by an element $\theta$   
of the group $H^2({\bf Z}^2, U(1))=U(1)$ of equivalence classes 
of 2-cocycles defined up to the coboundary of 1-cocycles: 
\be
U_1U_2=e^{2\pi i\theta}U_2U_1.  
\label{nctorus}
\ee
Then the compactified theory is obtained by restricting to the subspace 
of the coordinates $X$'s which are invariant under the ${\bf Z}^2_\theta$ 
action: 
\be
U_i^{-1}X^jU_i=X^j+2\pi \ell_s\delta^j_i,~~~i,j=1,2.
\label{quocond}
\ee
This condition is known as the quotient condition. 
Thus (\ref{quocond}) means that 
$$
X^i\in {\bf R}^2/{\bf Z}^2_{\theta},
$$
where ${\bf Z}^2_\theta$ denotes the projective representation 
(\ref{nctorus}). Note that (\ref{nctorus}) happens to define the algebra 
of functions 
${\cal A}(T^2_{\theta})$ on a noncommutative torus $T^2_{\theta}$.

In \cite{zumi}, a concrete representation for each $\theta$ is given 
by introducing a quantum plane coordinate 
\be
[\sigma_i, \sigma_j]=2\pi i\theta\epsilon_{ij}.
\label{qplane}
\ee
Then operators defined as  
\be
\tilde{U}_i=e^{i\sigma_i},
\label{dualgenerator}
\ee 
generate the algebra ${\cal A}(T^2_{-\theta})$ which is (one of) the 
dual algebra to ${\cal A}(T^2_{\theta})$ (see below): 
\be
\tU_1\tU_2=e^{-2\pi i\theta}\tU_2\tU_1.
\label{dualalg}
\ee
Using $\sigma_i$ and their partial derivative operators satisfying 
$$
[\partial_i, \sigma_j] = \delta_{ij}, 
$$
$$
[\partial_i, \partial_j] = 0,
$$
$U_i$ and $X^i$ can be expressed as 
\bea
U_i & = & e^{i\sigma_i}e^{2\pi\theta\epsilon_{ij}\partial_j}, \nono \\
X^i & = & q^i+A^i(\tU_i), \nono \\ 
q^i & = & -2\pi i\ell_s D_i,~~~~D_1=\partial_1,~~~
D_2=\partial_2-if\sigma_1,
\label{ncrep}
\eea
where $A^i$ is an arbitrary function of $\tU_i$. 
For later convenience, we have taken a representation 
in such a way that $q^i$'s are noncommutative: 
\be
[q^1,q^2]=-ik,~~~k\equiv -(2\pi\ell_s)^2f.
\label{defofkandf}
\ee 
In D-brane matrix models $q^i$ correspond to a classical solution 
to the equation of motion for $X^i$ and $A^i$ correspond to a fluctuation 
around $q^i$. 

Then the `dual lattice' corresponding to $T^2_{\theta}$ is given 
as the algebra of all operators that commute with $U_i$ \cite{cds}. 
In $\cite{ho,zumi}$, this algebra is represented as that of the sections 
on the twisted bundle of the adjoint representation of $U(n)$ 
on $T^2_{\theta}$ in the following way. 
First, introduce $n\times n$ matrices $U$, $V$ satisfying  
\be
UV=e^{-2\pi im/n}VU,
\ee
where $m$ is an integer. Without loss of generality, we assume that 
$n$ and $m$ are relatively prime. Taking an appropriate basis, such 
matrices are represented as 
\be
U_{kl}=\exp(2\pi ikm/n)\delta_{kl},~~~~~V_{kl}=\delta_{k+1,l}.
\label{uvrep}
\ee
Then the sections on adjoint bundles are generated by 
operator-valued matrices expressed as 
\be
Z_1=\exp\left(i\frac{\sigma_1}{n-m\theta}\right)V^b,~~~~~
Z_2=\exp\left(i\frac{\sigma_2}{n}\right)U^{-b},
\label{dualgen}
\ee
where $b$ is an integer satisfying $an-bm=1$ for some integer $a$. 
They indeed commute with the $U_i$'s and are also generators 
of the algebra of functions on a new noncommutative torus 
${\bf Z}^2_{\theta'}$ where 
\be
\theta'=\frac{a(-\theta) + b}{m(-\theta) +n}. 
\ee
Under this representation, the momentum lattice corresponding to 
(\ref{quocond}) is nothing other than ${\bf Z}^2_{\theta'}$. In fact, 
for any element $Z_i\in {\bf Z}^2_{\theta'}$, an operator $z_i$ 
defined by $e^{iz_i}=Z_i$ satisfies 
\be
\exp i\Tr(U_i^{-1}X^jU_iz_j)=\exp i\Tr(U_i^{-1}X^jz_jU_i)
=\exp i\Tr(X^jz_j),
\label{ncsingle}
\ee
in the analogy of (\ref{single}). Here $\Tr$ denotes the composition 
of the trace $\tr$ over $n\times n$ matrices and the integration over 
$\sigma_i$ $\cite{ho}$:  
\be
\Tr(f(Z_1,Z_2))=\int d\sigma_1d\sigma_2 \tr f(Z_1,Z_2).
\ee

As shown in \cite{tetal,ho,zumi}, in the context of Matrix theory 
unitary operators $U_i$ and scalar fields $X^i$ are represented 
as operators (\ref{ncrep}) for $k\neq 0$ acting on the twisted $U(n)$ 
fundamental bundle on the (non)commutative torus,  
\bea
\phi(\sigma_1+2\pi, \sigma_2) & =
 & \Omega_1(\sigma_1, \sigma_2)\phi(\sigma_1, \sigma_2),
\nono \\
\phi(\sigma_1, \sigma_2+2\pi) & =
 & \Omega_2(\sigma_1, \sigma_2)\phi(\sigma_1, \sigma_2), 
\label{twistedfundbdle}
\eea     
where $\Omega_i$ are transition functions 
\be
\Omega_1(\sigma_1, \sigma_2)=e^{i\alpha\sigma_2}U,~~~
\Omega_2(\sigma_1, \sigma_2)=V,
\label{transfun}
\ee
with $U$, $V$ given in (\ref{uvrep}) and $\alpha$ being a certain number 
determined below. 
They satisfy the consistency condition 
\be
\Omega_1(\sigma_1,\sigma_2+2\pi)\Omega_2(\sigma_1,\sigma_2)
=\Omega_2(\sigma_1+2\pi,\sigma_2)\Omega_1(\sigma_1,\sigma_2).
\label{consistency}
\ee
On the other hand, covariant derivatives which appear in (\ref{ncrep}) 
satisfy the twisted boundary conditions 
\bea
D_i(\sigma_1+2\pi,\sigma_2) & = & \Omega_1(\sigma_1,\sigma_2)
                                       D_i(\sigma_1,\sigma_2)
                                  \Omega_1(\sigma_1,\sigma_2)^{-1}, \nono \\
D_i(\sigma_1,\sigma_2+2\pi) & = & \Omega_2(\sigma_1,\sigma_2)
                                       D_i(\sigma_1,\sigma_2)
                                  \Omega_2(\sigma_1,\sigma_2)^{-1}. 
\label{twistedbc}
\eea 
Combined with (\ref{consistency}), this leads to the remarkable relation 
\be
\frac{1}{2\pi f}+\theta=\frac{n}{m}. 
\label{ncconsistency}
\ee 
This equation means that the magnetic flux, the noncommutativity 
of the base space torus, the rank of the gauge group and the number of the 
twist must be related with each other from the requirement of the gauge theory 
on the noncommutative torus, since $[D_1,D_2]=-if$ and $\alpha=m/n$.

\section{D-instantons on $T^2$ and $T^2_{\theta}$}
\setcounter{equation}{0}

In this section, we consider a boundary state together with operators 
acting on it corresponding to a configuration of D-instantons on $T^2$ 
and $T^2_{\theta}$ in the bosonic string theory. Then one can show that 
the target space noncommutativity is reflected by that of the momentum 
operators and hence noncommutative structures in the D-instanton world volume 
and the target space are dual to each other. In order to clarify the role of 
the noncommutativities, we proceed step by step starting from the
commutative case ($\theta=0, k=0$). 
  
\par\bigskip\noindent
\underline{$\theta=0, k=0$}
\par\medskip\noindent

Let us begin with defining a boundary state corresponding to D-instantons 
on commutative torus $T^2$. We propose that it takes the following form: 
\be
\sta{\sigma}=\sum_{w^i\in {\bf Z}}e^{i\sigma_iw^i}
             \ket{X^i=2\pi\ell_sw^i}_{-1},~~~~~
\sigma_i\in {\bf R}/2\pi {\bf Z},~~~~~i=1,2
\label{ymstate}
\ee 
where $\ket{X^i=2\pi\ell_sw^i}_{-1}$ denotes the Dirichlet boundary state 
defined as 
\be
\ket{X^i=2\pi\ell_sw^i}_{-1}=e^{-2\pi i\ell_sp_iw^i}\ket{X^i=0}_{-1}.
\label{Dstate}
\ee
Here we define $\ket{X^i=0}_{-1}$ to be the coherent state satisfying  
\be
X^i(s)\ket{X^i=0}_{-1}=0,
\ee
and $p_i$ is the center of mass momentum of the string. 
Since (\ref{Dstate}) labeled by $w_i\in {\bf Z}$ are all 
physically equivalent in the torus compactification, it is natural to consider 
the state $\sta{\sigma}$ given by a superposition of these states 
as in (\ref{ymstate}). Then it is easy to show that on $\sta{\sigma}$ 
the center of mass coordinate of the string $x^i$ (conjugate to $p_i$) becomes 
\be
x^i\sta{\sigma}=-2\pi i\ell_s\partial_i\sta{\sigma},
\label{BS_x}
\ee  
and, therefore, the coordinate operator of D-instantons $q^i$ is represented 
as    
\be
q^i=-2\pi i\ell_s\partial_i.
\label{Dinstcoord}
\ee
On the other hand, on $\sta{\sigma}$, $p_j$ satisfies%
\footnote{Since boundary state is a interaction vertex between string
and D-brane,
(\ref{BS_x}) and (\ref{momconserv}) are analogues of the relations
$x_1\delta(x_1-x_2)=x_2\delta(x_1-x_2)$ and
$-i(\partial_1 + \partial_2)\delta(x_1-x_2)=0$ satisfied by the local vertex
$\delta(x_1-x_2)$ for the point particles.}
\be
\left(p_j-\frac{\sigma_j}{2\pi\ell_s}\right)\sta{\sigma}=0, 
\label{momconserv}
\ee
since by definition, 
\be
e^{2\pi i\ell_sp_j}\ket{X^i=2\pi i\ell_sw^i}_{-1}
=\ket{X^i=2\pi\ell_s(w^i-\delta_i^j)}_{-1},
\ee
and hence 
\bea
e^{2\pi i\ell_sp_j}\sta{\sigma} & = & 
\sum_{w^i}e^{i\sigma_iw^i}\ket{X^i=2\pi\ell_s(w^i-\delta_i^j)}_{-1} \nono \\
 & = & \sum_{w^i}e^{i\sigma_i(w^i+\delta_i^j)}
       \ket{X^i=2\pi\ell_sw^i}_{-1} \nono \\
 & = & e^{i\sigma_j}\sta{\sigma} = \tU_j\sta{\sigma}.  
\eea
{}From (\ref{momconserv}), we can identify the `momentum' $\pi_j$ carried by 
the D-instanton configuration as 
\be
\pi_j=-\frac{\sigma_j}{2\pi\ell_s}, 
\label{Dinstmom}
\ee
because (\ref{momconserv}) represents the total momentum conservation.
Note that they indeed satisfy the Heisenberg algebra
$[ q^i , \pi_j ] = i \delta^i_j$.

Comparing (\ref{Dinstcoord}) and (\ref{Dinstmom}) with (\ref{dualgenerator}) 
and (\ref{ncrep}), we see that $\sta{\sigma}$ naturally realizes 
the representation corresponding to D-instantons on $T^2$ with $f=0$.  

{}From (\ref{ymstate}) $\sigma$ is a dual coordinate to the winding number. 
This is consistent with usual arguments in Matrix theory where matrices have
extra indices corresponding to the multiple cover of $S^1$ and the following
quotient relation is put:
\be
X^i_{m,n}=X^i_{m-1,n-1}+2\pi RI\delta_{mn}.
\ee
Then the Fourier transformation with respect to the extra indices leads to
dual coordinate $\sigma$ which gives the base space of the SYM description.
      
\par\bigskip\noindent
\underline{$\theta\neq 0, k=0$}
\par\medskip\noindent

Now let us turn to the string theory around D-instantons 
on the noncommutative torus, namely, $\theta\neq 0$. In this case, it is 
natural to define a state $\sta{\sigma}$ of the form (\ref{ymstate}) 
with $\sigma$ given by (\ref{qplane}). 
Note that $\sta{\sigma}$ defined as such is not the eigenstate of the 
abstract operator $\sigma$. It is easy to find that  
on $\sta{\sigma}$ $x^i$ is again given as 
\be
x^i\sta{\sigma} = q^i\sta{\sigma}, 
\label{ncDinstcoord}
\ee
with $q^i$ given as (\ref{Dinstcoord}), however, (\ref{momconserv}) is 
modified as 
\be
(p_j-\frac{\theta}{4\pi\alpha'}\epsilon_{jk}x^k-\frac{\sigma_j}{2\pi\ell_s})
\sta{\sigma}=0.  
\label{ncDinstmom}
\ee
In order to identify $\pi_j$, we propose that it should satisfy 
Heisenberg algebra with $q^i$ being (\ref{Dinstcoord}) 
and commute with $U_i$ according to the formulation developed in section 2. 
Then we find that $\pi_j$ is again expressed as (\ref{Dinstmom}). 
{}From (\ref{ncDinstmom}), we obtain the total momentum conservation 
in the following form: 
\be
(p_j-\frac{\theta}{4\pi\alpha'}\epsilon_{jk}x^k+\pi_j)\sta{\sigma}=0.  
\label{modifiedmom}
\ee
Note that the center of mass momentum of the string seems to become 
$p_j-\frac{\theta}{4\pi\alpha'}\epsilon_{jk}x^k$ in (\ref{modifiedmom}). 
This form is an analogue of the momentum of a charged particle moving 
in a magnetic field. 
In the next section, by considering a configuration of D-instantons on $T^2$ 
with a constant 2-form field flux $B_{ij}$, we will find it quite natural 
to make the identification $\theta=B_{12}$. 
Moreover, using (\ref{ncDinstcoord}), this equation can be rewritten as 
\be
(p_j-\frac{\theta}{4\pi\alpha'}\epsilon_{jk}q^k+\pi_j)\sta{\sigma}=0. 
\label{modifiedmom2} 
\ee
It states that in the case of $\theta\neq 0$ the momentum operator 
$\pi_j$ has an extent introduced by the second term in (\ref{modifiedmom2}) 
in the momentum space due to the target space noncommutativity and 
that as a consequence the center of mass momentum of the string $p_i$ 
satisfies the total momentum conservation in a modified form as above. 
It would be interesting to examine relation between this extension 
in the momentum space and the target space noncommutativity. 
In any case, the representation on $\sta{\sigma}$ with $\sigma$ 
given by (\ref{qplane}) turns out to correspond to a configuration 
of D-instantons on $T^2_{\theta}$, since $q^i$ is represented 
as (\ref{ncrep}) and $\pi_j$ is on the dual lattice. 
Remarkably, on $\sta{\sigma}$, the `momentum operators' of D-instantons 
do not commute with each other reflecting the noncommutativity 
of the target space torus:  
\be
[\pi_1, \pi_2]=\frac{1}{(2\pi\ell_s)^2}[\sigma_1,\sigma_2]
              =\frac{i}{2\pi\ell_s^2}\theta. 
\label{momnc}
\ee
Moreover, for any operator $W$ which depends on $\sigma_i$ and $\partial_i$, 
one can define corresponding matrix elements $W_{m^in^j}$ as 
\be
W\sta{\sigma}=\sum_{n^j}We^{i\sigma_jn^j}\ket{X^i=2\pi\ell_sn^j}_{-1}
             =\sum_{m^i,n^j}e^{i\sigma_im^i}W_{m^in^j}
              \ket{X^i=2\pi\ell_sn^j}_{-1}.
\ee
Then if one imposes the quotient condition (\ref{quocond}) on 
$X^i=q^i+A^i(\tU_i)$ with $q^i$ being the coordinate operator of D-instantons 
given in (\ref{Dinstcoord}), it can be rewritten as the conditions 
on matrix elements like  
\bea
q^j_{n^l+\delta_{il}~m^k+\delta_{ik}} 
              & = & q^j_{n^lm^k}+2\pi\ell_s\delta_{ij}\delta_{n^lm^k}, \nono \\
A^j_{n^l+\delta_{il}~m^k+\delta_{ik}} 
              & = & e^{-\pi i\theta\epsilon_{ik}(m^k-n^k)}A^j_{n^lm^k}.
\eea
Since $q^i$ is a diagonal matrix in this representation, these lead to 
\be
X^j_{n^l+\delta_{il}~m^k+\delta_{ik}}
=e^{-\pi i\theta\epsilon_{ik}(m^k-n^k)}
 (X^j_{n^lm^k}+2\pi\ell_s\delta_{ij}\delta_{n^lm^k}). 
\ee 
Under the identification $\theta=B_{12}$ which we made above, 
this is nothing other than the relation proposed in \cite{KO}. 
This confirms that in the present case it is natural to choose 
the representation on $\sta{\sigma}$. 
Thus we conclude that on $T^2_{\theta}$ D-instanton momentum operators have 
the noncommutative structure as in (\ref{momnc}), and this is also the case 
on $T^2$ with the constant 2-form field flux. 

As emphasized in the introduction, the world volume noncommutativity of 
D-instantons is characterized by that of the space-time coordinates of 
these D-instantons, while as shown in this section, 
the target space noncommutativity causes that of the momentum operators of 
D-instantons. In this sense we can say that they are dual to each other. 

For later convenience, it is useful to see that for $\theta\neq 0$, 
due to the ordering ambiguity, it is possible that $\sta{\sigma}$ 
takes a slightly different form from (\ref{ymstate})  
\be
\sta{\sigma}=\sum_{w^i\in {\bf Z}}e^{i\sigma_2w^2}e^{i\sigma_1w^1}
             \ket{X^i=2\pi\ell_sw^i}_{-1},
\label{ymstate2}
\ee
with $\sigma$ given by (\ref{qplane}). $q^i$ is expressed 
as (\ref{Dinstcoord}) on this state, while (\ref{modifiedmom}) turns into 
\be
(p_1-\frac{\theta}{2\pi\alpha'}x^2+\pi_1)\sta{\sigma}=0,~~~
(p_2+\pi_2)\sta{\sigma}=0,
\label{modifiedmom3}
\ee
where $\pi_i$ is the same as (\ref{Dinstmom}). 
The modified form (\ref{ymstate2}) is formally obtained from (\ref{ymstate})
by multiplying a factor $\exp(i{\theta\over 4\pi\alpha'}x^1x^2)$,
which may be interpreted as a gauge transformation for the 2-form gauge potential under the identification of $\theta = B_{12}$.

\par\bigskip\noindent
\underline{$\theta=0, k\neq 0$}
\par\medskip\noindent

Next let us discuss the case of $k\neq 0$. We first consider such a 
configuration on the commutative torus $T^2$. We propose that a boundary 
state corresponding to it is given as 
\be
\sta{\sigma_1,\sigma_2}=\sum_{w^i\in{\bf Z}}e^{i\sigma_iw^i}
                        \ket{2\pi\ell_s(w^1, w^2-f\sigma_1)}, 
\label{twistedymstate}
\ee
where $f=-k/(2\pi\ell_s)^2$, and 
\be
\ket{2\pi\ell_s(w^1, w^2-f\sigma_1)}\equiv 
\ket{(X^1,X^2)=2\pi\ell_s(w^1, w^2-f\sigma_1)}.  
\ee
In fact, it is easy to find that 
\bea
(x^1+(2\pi\ell_s)^2fp_2)\sta{\sigma_1,\sigma_2}
 & = & -2\pi i\ell_s\partial_{\sigma_1}\sta{\sigma_1,\sigma_2}, \nono \\
x^2\sta{\sigma_1,\sigma_2}
 & = & -2\pi i\ell_s(\partial_{\sigma_2}-if\sigma_1)\sta{\sigma_1,\sigma_2},
\label{Dstcoord}
\eea
from which we read the coordinates of D-instantons $q^i$ as  
\be
q^1=-2\pi i\ell_s\partial_{\sigma_1},~~~
q^2=-2\pi i\ell_s(\partial_{\sigma_2}-if\sigma_1).
\ee
These imply that $\sta{\sigma_1,\sigma_2}$ exactly provides 
the representation given in (\ref{ncrep}). 
Note that in (\ref{Dstcoord}) $x^1$ does not quite coincide with $q^1$. 
We regard it as a signal that for $k\neq 0$ each D-instanton is spreading 
in size $\sim k$ and that strings are attached to such an extended object.%
\footnote{It may also suggest that for $k\neq 0$ a combination $x^1-kp_2$ 
plays a role of a new string coordinate 
in which the center of mass coordinates of the string itself become 
noncommutative. A similar result is also obtained in \cite{iran,chuho} 
by considering the quantization of open strings ending on D-branes. The
noncommutativity of string coordinates already appeared in earlier works
\cite{ikks,sakamoto}.}
In order to put a more physical interpretation on what happens, 
we analyze this configuration by means of another method in the next section. 
Also note that 
\be
e^{2\pi i\ell_sp^i}\sta{\sigma_1,\sigma_2} 
    = e^{i\sigma_i}\sta{\sigma_i,\sigma_2}, 
\ee
and, therefore, we obtain 
\be
\pi_i=-\frac{\sigma_i}{2\pi\ell_s}.
\ee
This is again consistent with the formulation in section 2. 
Moreover, one can generalize (\ref{twistedymstate}) to include a index 
$r=1,\cdots,n$ corresponding to the Chan-Paton index in the open string sector 
in the following way:
\be
\sta{\sigma_1,\sigma_2}^r=\sum_{w^i\in{\bf Z}}
                          e^{i\sigma_1w^1+i(\sigma_2+2\pi r)w^2/n}
                          \ket{2\pi\ell_s(w^1, w^2/n-f\sigma_1)}, 
\label{CPtwistedymstate}
\ee
which coincides with (\ref{twistedymstate}) for $n=1$. 
Like (\ref{twistedymstate}), $\sta{\sigma_1,\sigma_2}^r$ satisfies 
\bea
(x^1+(2\pi\ell_s)^2fp_2)\sta{\sigma_1,\sigma_2}^r
 & = & -2\pi i\ell_s\partial_{\sigma_1}\sta{\sigma_1,\sigma_2}^r, \nono \\
x^2\sta{\sigma_1,\sigma_2}^r 
 & = & -2\pi i\ell_s(\partial_{\sigma_2}-if\sigma_1)\sta{\sigma_1,\sigma_2}^r,
\nono \\
e^{2\pi i\ell_sp^i}\sta{\sigma_1,\sigma_2}^r 
 & = & e^{i\sigma_i}\sta{\sigma_i,\sigma_2}^r. 
\eea
Thus on $\sta{\sigma_1,\sigma_2}^r$ the coordinates of D-instantons $q^i$ 
are represented as $n\times n$ matrices   
\be
q^1=-2\pi i\ell_s\partial_{\sigma_1}I,~~~
q^2=-2\pi i\ell_s(\partial_{\sigma_2}-if\sigma_1)I,
\label{q^irep}
\ee
and $\pi_i$ are represented as 
\be
\pi_i=-\frac{\sigma_i}{2\pi\ell_s}I,
\ee
as expected. Recalling that as commented in section 2, representations 
like (\ref{q^irep}) are given on the twisted $U(n)$ fundamental bundle 
on $T^2$ in the context of Matrix theory, we expect that the state 
(\ref{CPtwistedymstate}) in string theory has structure 
similar to the twisted $U(n)$ fundamental bundle. 
In fact, (\ref{CPtwistedymstate}) satisfies the same twisted boundary 
condition as in (\ref{twistedfundbdle}), 
\bea
\sta{\sigma_1+2\pi,\sigma_2}^r
  & = & e^{im/n\sigma_2}U_{rs}\sta{\sigma_1,\sigma_2}^s, \nono \\
\sta{\sigma_1,\sigma_2+2\pi}^r 
  & = & V_{rs}\sta{\sigma_1,\sigma_2}^s,
\label{bcofsigma}
\eea
provided that 
\be
\frac{1}{2\pi f}=\frac{n}{m},
\label{relation}
\ee
which is just the equation obtained by setting $\theta=0$ 
in (\ref{ncconsistency}). Later we will discuss this point 
including the case of $\theta\neq 0$. 

\par\bigskip\noindent
\underline{$\theta\neq 0, k\neq 0$}
\par\medskip\noindent

Finally let us discuss the case of $\theta\neq 0$, $k\neq 0$. Using the quantum 
plane coordinate (\ref{qplane}), a boundary state corresponding to 
a configuration with $\theta\neq 0, k\neq 0$ including the index 
$r=1,\cdots,n$ is given as 
\be
\sta{\sigma_1,\sigma_2}^r
=\sum_{w^i\in{\bf Z}}e^{i(\sigma_2+2\pi r)w^2/n}e^{\sigma_1w^1}
 \ket{2\pi\ell_s(w^1, (1+2\pi f\theta)w^2/n-f\sigma_1)}. 
\label{NCtwistedymstate}
\ee
Note that this state reproduces (\ref{CPtwistedymstate}) for $\theta=0$. 
Little examination shows that this state satisfies the following equations: 
\bea
(x^1+(2\pi\ell_s)^2fp_2)\sta{\sigma_1,\sigma_2}^r
 & = & -2\pi i\ell_s\partial_{\sigma_1}\sta{\sigma_1,\sigma_2}^r, 
\label{x1}\\ 
x^2\sta{\sigma_1,\sigma_2}^r 
 & = & -2\pi i\ell_s(\partial_{\sigma_2}-if\sigma_1)\sta{\sigma_1,\sigma_2}^r,
\label{x2} \\
e^{2\pi i\ell_s(p_1-\frac{\theta}{2\pi\alpha'}x^2)}\sta{\sigma_1,\sigma_2}^r
 & = & e^{i(1+2\pi f\theta)\sigma_1}\sta{\sigma_1,\sigma_2}^r,
\label{p1} \\
e^{i(1+2\pi f\theta)2\pi\ell_sp_2}\sta{\sigma_1,\sigma_2}^r
 & = & e^{i\sigma_2}\sta{\sigma_1,\sigma_2}^r.
\label{p2}
\eea
{}From (\ref{x1}) and (\ref{x2}), on $\sta{\sigma_1,\sigma_2}^r$ 
the coordinates of D-instantons are given as  
\bea
q^1 & = & -2\pi i\ell_s\partial_{\sigma_1}I, \label{q1} \\
q^2 & = & -2\pi i\ell_s(\partial_{\sigma_2}-if\sigma_1)I.\label{q2}
\eea
Next let us identify the momenta $\pi_i$ carried by the D-instanton 
configuration. As for $\pi_2$, combining (\ref{p2}) with the total momentum 
conservation 
\be
(p_2+\pi_2)\sta{\sigma_1,\sigma_2}^r=0,
\ee
we obtain 
\be
\pi_2=-\frac{\sigma_2}{(1+2\pi f\theta)2\pi\ell_s}I.
\label{pi2}
\ee
In fact, we see that (\ref{pi2}) is consistent with both Heisenberg algebra 
and formulation in section 2 because $q^2$ is represented as in (\ref{q2}), 
and (\ref{pi2}) commutes with $U_i$. As for $\pi_1$, however, 
if we require that $\pi_1$ should satisfy the Heisenberg algebra 
with $q^1$ given by (\ref{q1}) and commute with $U_i$, it must be given as 
\be
\pi_1=-\frac{\sigma_1}{2\pi\ell_s}I.
\label{pi1}
\ee
Compared with (\ref{p1}), this equation implies that 
the total momentum conservation in the 1-direction is modified 
in the following form: 
\be
\left(p_1-\frac{\theta}{2\pi\alpha'}x^2+(1+2\pi f\theta)\pi_1\right)
\sta{\sigma_1,\sigma_2}^r=0, 
\label{twistedmomconserv}
\ee
or 
\be
\left(p_1-\frac{\theta}{2\pi\alpha'}q^2+(1+2\pi f\theta)\pi_1\right)
\sta{\sigma_1,\sigma_2}^r=0. 
\ee
These equations admit of the same interpretation as in (\ref{modifiedmom}) 
and (\ref{modifiedmom2}). Note that (\ref{twistedmomconserv}) 
agrees with the first equation in (\ref{modifiedmom3}) for $f=0$. 
Thus we have verified that $\sta{\sigma_1,\sigma_2}^r$ indeed 
corresponds to the configuration in the case of $\theta\neq 0$, $k\neq 0$.  
  
As emphasized in (\ref{bcofsigma}), whether $\theta=0$ or not, 
our state $\sta{\sigma_1,\sigma_2}^r$ in the case of $k\neq 0$ 
satisfies the same twisted boundary condition as in (\ref{twistedfundbdle}), 
\bea
\sta{\sigma_1+2\pi,\sigma_2}^r
  & = & e^{im/n\sigma_2}U_{rs}\sta{\sigma_1,\sigma_2}^s, \nono \\
\sta{\sigma_1,\sigma_2+2\pi}^r 
  & = & V_{rs}\sta{\sigma_1,\sigma_2}^s,
\eea
provided that 
\be
\frac{1}{2\pi f}+\theta=\frac{n}{m},
\label{generalrelation}
\ee
which is just the equation obtained in (\ref{ncconsistency}). 
A similar equation is also obtained in \cite{cds}. It is worth noticing 
that a boundary state which corresponds to the twisted fundamental bundle 
on the (non)commutative torus can be constructed 
in the context of string theory, and that it reproduces the same condition 
as that required from a consistency of the twisted bundle.  
Although at present this correspondence is formal, it is expected that 
string theory around a D-instanton configuration with $\theta\neq 0$, 
$k\neq 0$ has certain structure described by the (fundamental, 
or possibly adjoint) twisted bundle on the noncommutative torus. 
For example, if (\ref{generalrelation}) is satisfied, it is easy to 
show that 
\bea
e^{2\pi i\ell_s(p_1-\frac{\theta}{2\pi\alpha'}x^2)}\sta{\sigma_1,\sigma_2}^r
 & = & (Z_1^n)_{rs}\sta{\sigma_1,\sigma_2}^s,  \\
e^{i\frac{2\pi\ell_s}{n-m\theta}p_2}\sta{\sigma_1,\sigma_2}^r
 & = & (Z_2)_{rs}\sta{\sigma_1,\sigma_2}^s. 
\eea
We see that operators $Z_i$ defined in (\ref{dualgen}) naturally appear 
in the right-hand sides of both equations. It may suggest that 
the statement mentioned above is indeed the case. 
Among others, it would be important to give a physical meaning to 
(\ref{generalrelation}) in the context of string theory. For this purpose, 
in the next section we consider a boundary state corresponding to 
a configuration of D-instantons on $T^2$ with a constant $B_{ij}$ flux 
and show that the relation (\ref{generalrelation}) is automatically satisfied 
with suitable identifications.

\section{D-instantons on $T^2$ with the 2-Form Field Flux}
\setcounter{equation}{0}

In this section, we consider a configuration of D-instantons on $T^2$ 
of size $2\pi\ell_s$ with the constant 2-form field $B_{ij}$ flux. 
According to the boundary state analysis made in \cite{ishi}, 
a configuration of D-instantons whose space-time coordinates are 
noncommutative can be regarded as that of a single D-string. 
We closely follow this approach and clarify the relationship 
between the noncommutativity of the world volume and the $B_{ij}$ flux. 

A configuration of D-instantons on $T^2$ we would like to consider is 
given by 
\be
[q^1,q^2]=-ik\neq 0,  
\label{dinstback}
\ee 
where $q^i$ are $n\times n$ matrices corresponding to the space-time 
coordinates of D-instantons, entries of which are in general noncommutative 
operators. In compactified Matrix theory, such matrices are explicitly 
constructed for finite $n$ \cite{tetal}. 
Note that the representation (\ref{ncrep}) of $q^i$ given 
in section 2 shows that $k$ is in proportion to the magnetic flux $f$ as   
\be
k=-(2\pi\ell_s)^2f,~~~~[D_1,D_2]=-if. 
\ee 
As stated in the introduction, $k$ characterizes the noncommutative structure 
of the world volume of D-instantons. 

The boundary state for the configuration (\ref{dinstback}) is 
\be
\ket{B}_{-1}=\tr P\exp\left(-i\int_0^{2\pi}ds P_i(s)q^i
                      \right)\ket{X^i=0}_{-1},
\ee 
where 
$P_i(s)$ is the canonical momentum of the string 
in the presence of the background 2-form 
\be
P_i(s)=\frac{1}{2\pi\alpha'}(G_{ij}\dot{X}^j-B_{ij}X'^j).
\ee

Then similarly to \cite{ishi}, this boundary state can be expressed 
as the path integral representation  
\be
\ket{B}_{-1}=\int [dq^1dq^2]\exp\left(\frac{i}{k}
             \int_0^{2\pi}ds q^1(s)\partial_sq^2(s)
             -i\int_0^{2\pi}ds P_i(s)q^i(s)
                                \right)\ket{X^i=0}_{-1}.
\ee
{}From trivial identities 
\be
0=\int [dq^1dq^2]\frac{\delta}{\delta q^j(s)}
                 \exp\left(\frac{i}{k}
                 \int_0^{2\pi}ds q^1(s)\partial_sq^2(s)
                -i\int_0^{2\pi}ds P_i(s)q^i(s)
                     \right)\ket{X^i=0}_{-1}, 
\ee
one obtains 
\be
\left[P_i(s)-\frac{1}{k}\epsilon_{ij}\partial_sX^j
\right]\ket{B}_{-1}
=0,
\ee
and, therefore, 
\bea
\lefteqn{0 = \left[\frac{1}{2\pi\ell_s}
        (\hat{n}_i-\frac{2\pi\alpha'}{k}\epsilon_{ij}\hat{w}^j)
             \right.} \nono \\
 & + & \left.\frac{1}{2\pi}\sqrt{\frac{1}{2\alpha'}}
            \sum_{n\neq 0}
            \left((G_{ij}+B_{ij}+\frac{2\pi\alpha'}{k}\epsilon_{ij})\alpha_n^j
                 +(G_{ij}-B_{ij}-\frac{2\pi\alpha'}{k}\epsilon_{ij})
                                                             \tilde{\alpha}_n^j
            \right)e^{ins}
      \right] \nono \\
 &   &  \times\ket{B}_{-1},
\label{Dinstbc}
\eea
where $G_{ij}$ is the metric of the $T^2$ and $\hat{n}_i$, $\hat{w}^j$ are 
momentum number and winding number operator, respectively. 
On the other hand, the boundary state $\ket{B}_1$ corresponding to 
a configuration of a D-string on $T^2$ of the metric $G'_{ij}$ 
with the background $B'_{ij}$, $F'_{ij}$ satisfies the following equation: 
\bea
\lefteqn{0 = \left[P_i(s)-F'_{ij}\partial_sX^j(s)
        \right]\ket{B}_1} \nono \\
 & = &  \left[\frac{1}{2\pi\ell_s}
       (\hat{n}'_i-2\pi\alpha' F'_{ij}\hat{w}'^j)\right. \nono \\
 & + & \left.\frac{1}{2\pi}\sqrt{\frac{1}{2\alpha'}}
             \sum_{n\neq 0}
             \left((G'_{ij}+B'_{ij}+2\pi\alpha' F'_{ij})\alpha_n^j
                  +(G'_{ij}-B'_{ij}-2\pi\alpha' F'_{ij})\tilde{\alpha}_n^j
             \right)e^{ins}
        \right] \nono \\
 &   & \times\ket{B}_1. 
\label{Dstbc}
\eea
Comparing (\ref{Dinstbc}) with (\ref{Dstbc}) shows that the configuration 
of $n$ D-instantons corresponding to (\ref{dinstback}) on $T^2$ with 
the $B_{ij}$ background coincides with that of a D-string on $T^2$ 
with the background 
\bea
\hat{n}'_i-2\pi\alpha' F'_{ij}\hat{w}'^j 
  & = & \hat{n}_i-\frac{2\pi\alpha'}{k}\epsilon_{ij}\hat{w}^j, 
\label{zeromode} \\
G'_{ij}+B'_{ij}+2\pi\alpha' F'_{ij}
 & = & G_{ij}+B_{ij}+\frac{2\pi\alpha'}{k}\epsilon_{ij}, \\
G'_{ij}-B'_{ij}-2\pi\alpha' F'_{ij}
  & = & G_{ij}-B_{ij}-\frac{2\pi\alpha'}{k}\epsilon_{ij}. 
\label{dstback}
\eea
{}From the second and third equation, we obtain 
\be
G_{ij}=G'_{ij},~~~
B_{ij}+\frac{2\pi\alpha'}{k}\epsilon_{ij}=B'_{ij}+2\pi\alpha' F'_{ij},
\label{wvandts}
\ee
and then `natural identification' is   
\be
B'_{ij}=B_{ij},~~~F'_{ij}=\frac{1}{k}\epsilon_{ij}. 
\label{spiden}
\ee 
The last equation is exactly the same as in \cite{ishi}. However, recalling 
that $B'_{ij}$ and $F'_{ij}$ are transformed to each other by the gauge 
transformations from the point of view of the D-string, we are tempted  
to consider that there exist some `gauge transformations' between $1/k$ 
and $B_{ij}$ and interpret the identification (\ref{spiden}) 
as a particular gauge fixing.     

Next let us address the problem of deriving the equation like 
(\ref{ncconsistency}) in the D-instantons configuration under consideration. 
If the $U(n)$ gauge theory on the noncommutative tours correctly describes 
the $n$ D-instantons on the torus with the $B_{ij}$ flux as proposed in 
\cite{dh,ho,zumi}, the consistency condition (\ref{ncconsistency}) must be 
automatically satisfied in the D-instanton dynamics. As we have shown above, 
we can regard this configuration as the D-string configuration on $T^2$ 
with the background 
$2\pi\alpha' {\cal F}'_{ij}\equiv B'_{ij}+2\pi\alpha' F'_{ij}$. 
Then by construction, the D-string have the D-instanton charge $n$ 
and hence 
\be
\frac{1}{2\pi}\int_{\mbox{\small\rm D-string world volume}}
{\cal F}'=n,~~~{\cal F}'\equiv {\cal F}'_{12}.  
\ee 
Assuming that it winds $m$ times around $T^2$, 
\be
\frac{1}{2\pi}\int_{T^2}{\cal F}'=\frac{n}{m}. 
\ee
According to the second equation in (\ref{wvandts}), this can be reinterpret 
in terms of the background of the configuration of D-instantons as 
\be
\frac{2\pi\alpha'}{k}+B_{12}=\frac{n}{m}.
\label{Dinstwvandts}
\ee  
Making the identification $B_{12}=\theta$ exactly yields 
(\ref{generalrelation}) 
\be
\frac{1}{2\pi f}+\theta=\frac{n}{m}, 
\ee 
as expected, up to the sign which can be absorbed by the trivial redefinition 
of the variables. Since $k$ and $\theta$ denote the noncommutativities of 
the world volume of D-instantons and the target space respectively, 
this result implies that they are closely related through the non-trivial 
equation (\ref{Dinstwvandts}), which can be naturally understood 
from the point of view of the twisted bundle on the noncommutative torus. 
Thus we confirm the proposal made in \cite{dh,ho,zumi} that 
$n$ D-instantons on the $T^2$ with $B_{ij}$ flux can be described 
by the $U(n)$ twisted gauge theory on the noncommutative torus 
by regarding the configuration of the D-instantons as that of a D-string 
in the context of the perturbative bosonic string theory. 
There $m$ is interpreted as the winding number of the resulting D-string. 
The same interpretation is also made in \cite{tetal,iran,ho,zumi,verlinde} 
and seems natural for the following reason, for example. Setting 
$\theta=B_{12}=0$ in (\ref{Dinstwvandts}) yields 
\be
2\pi kn=(2\pi\ell_s)^2m,
\ee
which is consistent with the physical picture that each D-instanton has 
`cell' of area $2\pi k$ because of the world volume noncommutativity 
(\ref{dinstback}) and the D-string wrapped $m$ times over $T^2$ consists of 
$n$ such cells.

\section{Conclusions and Discussions}

We have constructed the boundary state which realizes the NCG
representations of D-branes on noncommutative torus. {}From our
construction the dual nature of the noncommutativity of the world
volume and that of the target space can be seen naturally.
Also our analysis is consistent with the conjecture that the gauge
theories on the world volumes of D-branes compactified on a torus
with the background 2-form field should be described by those on the
noncommutative torus.

Some remarks are in order:

We have used a particular representation of the D-brane coordinates $q^i$
\be
q^1=-2\pi i\ell_s\partial_{\sigma_1},~~~
q^2=-2\pi i\ell_s(\partial_{\sigma_2}-if\sigma_1).
\ee
There is an arbitrariness in the term proportional to $f$ due to a gauge transformation by an arbitrary function of $\sigma_j$.
It is an automorphism of the algebra on the noncommutative world volume
because it preserves the algebra of $q^i$'s and $\pi_j$'s.
Thus, in the sense of NCG, it should give (a part of) diffeomorphism of
the world volume. Actually, a function of $\sigma_j$ is
nothing but a function of $\pi_j$ so that it brings about a canonical
transformation of $q^i$. This is also consistent with the discussions in
\cite{ishi} that the longitudinal translation of the D-brane corresponds to
a kind of gauge transformation in the dual picture.

On the other hand, as noted below (\ref{modifiedmom3}),
one can modify the representation of the boundary state by a function of
$x^i$ which changes the expression of momentum conservation.
Under the identification of the parameter $\theta$ with 2-form gauge potential,
this transformation effectively gives gauge transformation of $B_{ij}$.
{}From the string point of view, the gauge transformation of
$B_{ij}$ is related
to the conformal transformation.
In this sense, if noncommutative structures discussed here
provide the space-time uncertainties 
\cite{yoneya,stur,jeviyone} which are considered to reflect 
the conformal symmetries both in the world volume and the target space, 
then our result seems to be consistent with the proposal in \cite{jeviyone} 
that two conformal symmetries play dual roles.
It will be interesting to consider an interpretation of the relation
(\ref{generalrelation}) in the light of the space-time uncertainty.

\section*{Acknowledgements}

T.K. would like to thank N. Ishibashi for valuable discussions 
and helpful comments.  
The work of M.K. was supported in part by the Sumitomo Foundation and the Grant-in-Aid for Scientific Research from the Ministry of Education, Science and Culture.

\end{document}